\documentclass[twocolumn, prl]{revtex4}

\usepackage{graphicx}
\usepackage{dcolumn}
\usepackage{bm}

\begin{document}

\title{ Mott Transition in Quasi-One-Dimensional Systems}

\author { S. Moukouri, E. Eidelstein}

\affiliation{ Racah Institute of Physics, Hebrew University, Jerusalem 91904 
Israel} 

\begin{abstract}
We report the application of the density-matrix renormalization group
 method to a spatially anisotropic two-dimensional Hubbard model at 
half-filling. We find a deconfinement transition induced by the 
transverse hopping parameter $t_y$ from an insulator to a  metal. 
 Therefore, if $t_y$ is fixed
in the metallic phase, increasing the interaction $U$ leads to a 
metal-to-insulator transition at a finite critical $U$. 
This is in contrast to the weak-coupling Hartree-Fock theory which 
predicts a nesting induced antiferromagnetic insulator for any $U>0$.  
\end{abstract}

\maketitle

The metal-insulator transition (MIT), also called the Mott 
transition \cite{mott}, is certainly one of the  most difficult challenge 
facing condensed matter theorists. Hubbard \cite{hubbard}, in a pioneering 
work, introduced a simple one-band Hamiltonian which has only two 
parameters, $t$ for the kinetic energy of the electrons and $U$ for 
the local electron-electron interactions. This model is at half-filling 
the model of reference for the MIT. In $D=1$, Lieb and Wu \cite{lieb-wu} 
obtained an exact solution by  using the Bethe ansatz. The ground state is 
an insulator for any $U/W >0$, where $W$ is the band width. Thus, the MIT 
occurs at the critical value $(U/W)_c=0$. In infinite dimensions, the 
dynamical mean-field theory (DMFT) \cite{metzner-vollhardt,georges-kotliar} 
predicts a critical point at $(U/W)_c \approx 1$. 

The discovery of layered materials, where the motion of electrons driving the
low energy physics is mostly confined in the layers, has raised
great interest into the two-dimensional (2D) Hubbard model. 
The physics at large $U/W \agt 1$ is now understood, the charge excitations 
are gapped, the spin excitation are described by the Heisenberg Hamiltonian 
which has long-range order (LRO) at $T=0$. But for $U/W \alt 1$, the physics
is still unclear. Our current knowledge about the weak-coupling regime 
is mostly drawn from the Hartree-Fock approximation and from  quantum Monte 
Carlo (QMC) simulations \cite{hirsch,varney}. The QMC results agree 
qualitatively with the Hartree-Fock prediction that the ground state is a 
Slater insulator for any $U/W > 0$. However, in most recent studies such 
as in Ref.\cite{varney}, even though considerable progress has been achieved 
in reaching larger systems, in the weak $U$ regime where the eventual gap is 
small, reliable extrapolations of the QMC data remain difficult to
achieve. It would thus be preferable to apply finite size scaling 
for data analysis instead of relying on extrapolations. 

More recently, extensions of the DMFT which include non-local fluctuations, 
the dynamical cluster approximation (DCA) \cite{moukouri-jarrell} or the 
cellular DMFT \cite{imada,park-haule}, have been applied to the 2D Hubbard 
model. The focus in these studies have mostly been to discuss the nature of 
the MIT within the paramagnetic solution of the DMFT equations. A systematic 
comparison of the possible ordered or disordered ground states as function 
of the cluster sizes is still lacking. Therefore, the issue as to whether 
or not quantum fluctuations destroy the Hartree-Fock solution in the 
half-filled 2D Hubbard model in the small $U$ regime remains open.

In this letter, we show that insight into this problem can be gained 
by studying the quasi-1D Hubbard model. We apply the two-step
density-matrix renormalization group (DMRG) \cite{moukouri}
to an array of coupled Hubbard chains. We find that there is a deconfinement 
transition from the 1D insulator toward a  metallic phase as the transverse 
hopping $t_y$ is increased from $0$. Hence, for a fixed $t_y$ in the metallic 
phase, there is a quantum phase transition (QPT) at a finite quantum 
critical point (QCP)$(U/W)_c$. This suggests, as seen in the limit of infinite 
dimensions \cite{metzner-vollhardt}, that by freezing out the local time 
dynamics, the Hartree-Fock approximation is unable to account for the 
physics of the Hubbard model even for weak interactions. We note that 
Biermann and coworkers \cite{biermann}  applied the chain-DMFT to the quasi-1D 
Hubbard model. They also found a deconfinement transition. However, they were 
restricted to paramagnetic solutions. Hence, unlike our study, they could 
not tell whether or not their metallic solution is the true ground state 
of the quasi-1D Hubbard model.

\begin{figure}
\begin{center}
$\begin{array}{c@{\hspace{0.25in}}c}
\includegraphics[width=2.75cm, height=2.75cm]{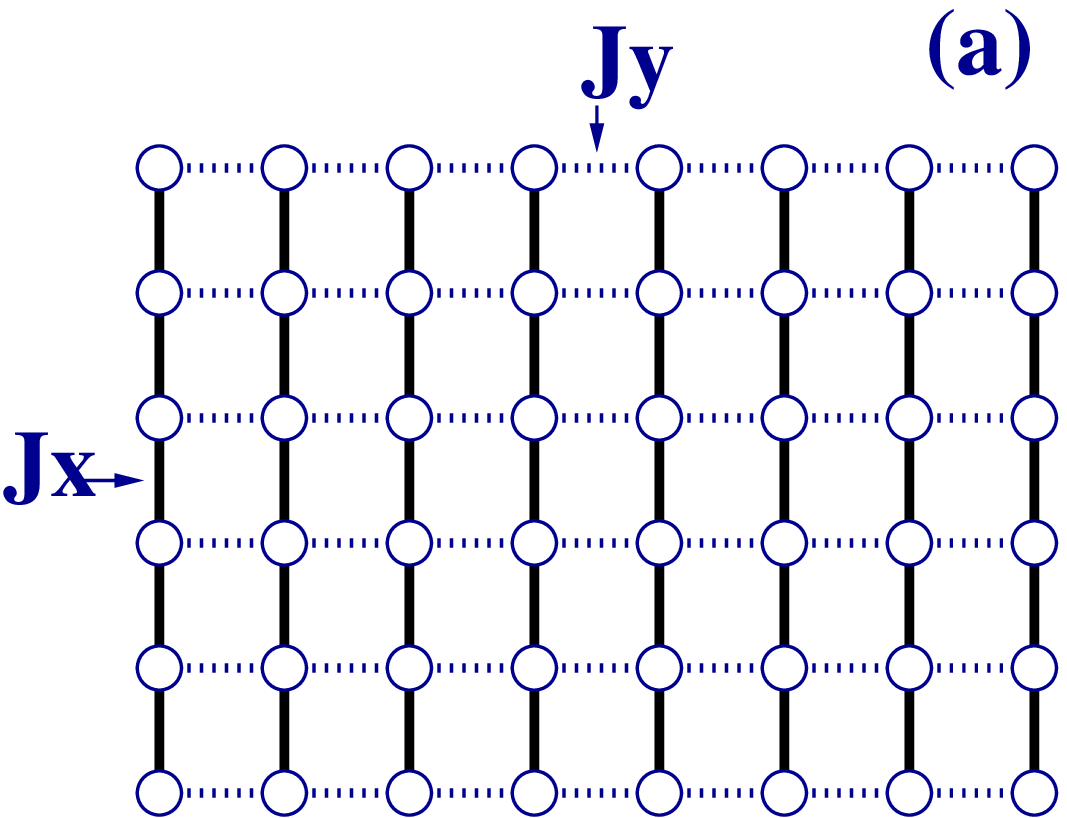}
\hspace{0.125cm}
\includegraphics[width=2.75cm, height=2.75cm]{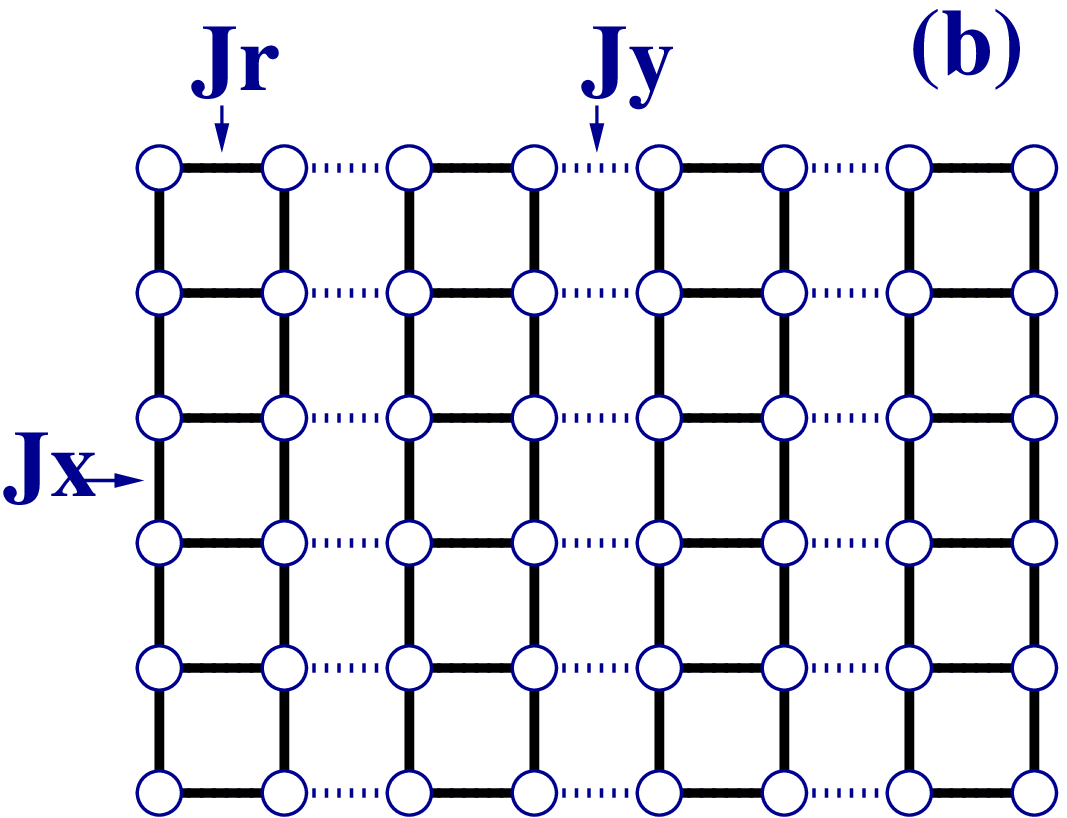}
\hspace{0.125cm}
\includegraphics[width=2.75cm, height=2.75cm]{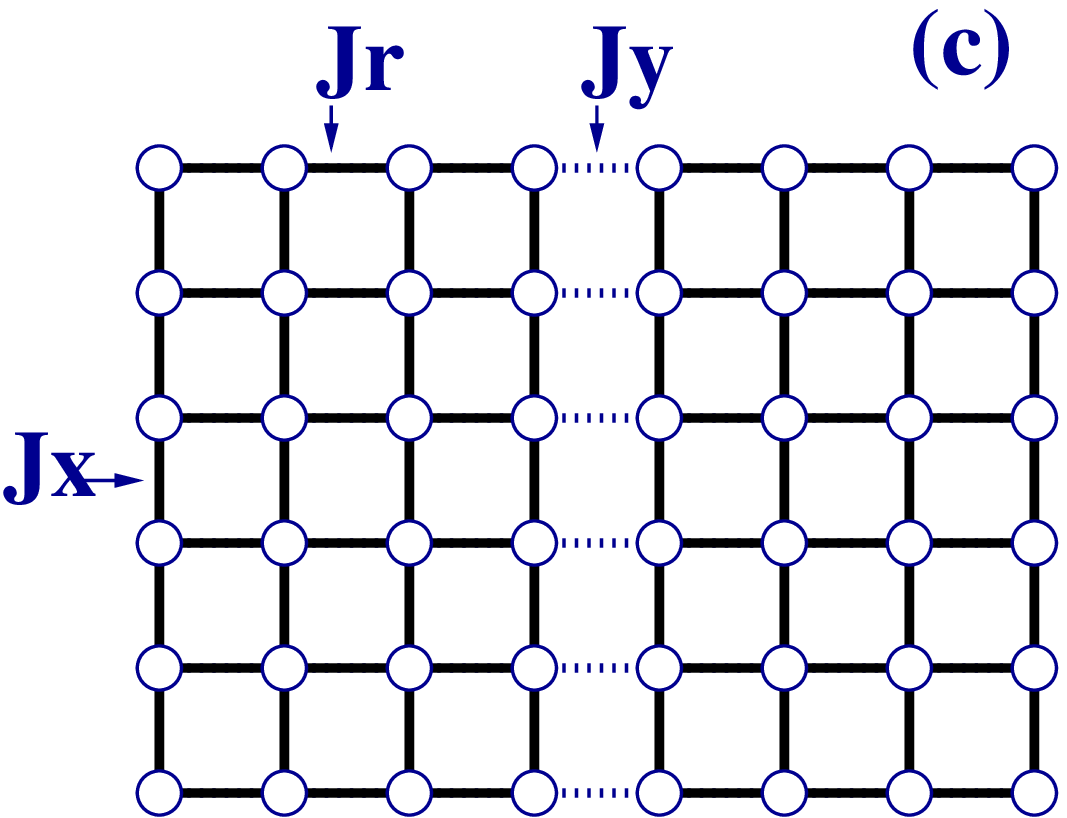}
\end{array}$
\end{center}
\caption{Clusters used as building blocks for the 2D square lattice:
  (a) chains; (b) two-leg ladders; (c) four-leg ladders:
$J_x=1$ is the coupling along the legs; $J_y$ is the inter-cluster
coupling, and $J_r$ is the coupling between the rungs for the ladders.}
\label{clusters}
\end{figure}

Let us  briefly describe the two-step DMRG method  introduced for
coupled chains in Ref.\cite{moukouri}. This method is generalized here for 
coupled multi-leg ladders. For a system of spins or electrons on an 
anisotropic square lattice of dimensions $L_x \times L_y$, the 
Hamiltonian may be written as:
$H=H_{intra}+H_{inter}=\sum_l h_l+g{\tilde H}_{inter}$,
 where $h_l$ are the Hamiltonians of 1D systems; $g\ll 1$
is the transverse coupling. ${\tilde H}_{inter}$ is of the same
magnitude as $H_{intra}$. $h_l$ can represent a single chain, a two-leg 
ladder or even a multi-leg ladder as illustrated in Fig.\ref{clusters} 
for a spin system. The DMRG analysis of $H$ is done in two steps. 
In the first step, the $m_2$ lowest eigenfunctions $\phi_n$ and 
eigenvalues $E_n$ from different charge-spin sectors of the 1D  
Hamiltonian $h_l$ are obtained by applying the conventional 1D DMRG 
algorithm \cite{white}. During this step, $m_1$ states are kept such 
that the lowest $m_2$ states are accurately 
computed. In the second step, $H$ is projected onto the tensor product of 
the $\phi_n$'s. Since the resulting effective 2D Hamiltonian is now 1D in the 
transverse direction, it can be studied using the conventional DMRG.  
Let for a given size, $E_0$ and $E_{m_2}$ be the lowest and the highest
DMRG energies kept for $h_l$. If $g \ll \Delta E=E_{m_2}-E_0$ ( in practice we
set $\frac{g}{\Delta E} \alt 10$), the two-step 
DMRG retains high accuracy. But as $L_x$ increases, $\Delta E$ decreases,
for a fixed $g$, it would be impossible to retain accuracy for arbitrary 
large $L_x$. Thus, successfully performing finite size analysis for the $g$ 
induced QPT will depend on the magnitude of the critical value of $g$ and 
on the density of the 1D spectrum. 

In order to avoid uncertainties related to extrapolations in the
regime where the gap is very small, it is preferable to perform finite 
size scaling analysis. In the vicinity of the QCP the gap, 
$\Delta(g) \propto (g_c-g)^{\nu}$, where $\nu$ is the correlation length 
exponent ( Lorentz invariance is assumed), and $g_c$ is the value of the 
$g$ at the QCP. The product $L \Delta$, where $L$ is the linear size of the 
system is given by a universal function, $L\Delta=f(C(g_c-g)L^{1/\nu})$ 
\cite{barber}, where $C$ is independent of $L$. At the critical point, 
$L \Delta=f(0)$ is independent of $L$. All the curves $\Delta (g)$ for 
different sizes should converge at $g=g_c$. In practice however there are 
finite size effects. It is necessary to extrapolate the different crossing 
points in order to precisely locate $g_c$.
Before studying the MIT in the Hubbard model, we first apply 
the two-step technique  to the QPT in quantum antiferromagnets (AFM). 
There are two motivations for this prelude on spin systems. First, well
controlled QMC are available, this will allow to gauge the performance
of the two-step DMRG. Second, the energy scales of the different systems
roughly span those of the Hubbard model when the interaction is varied.

\begin{figure}
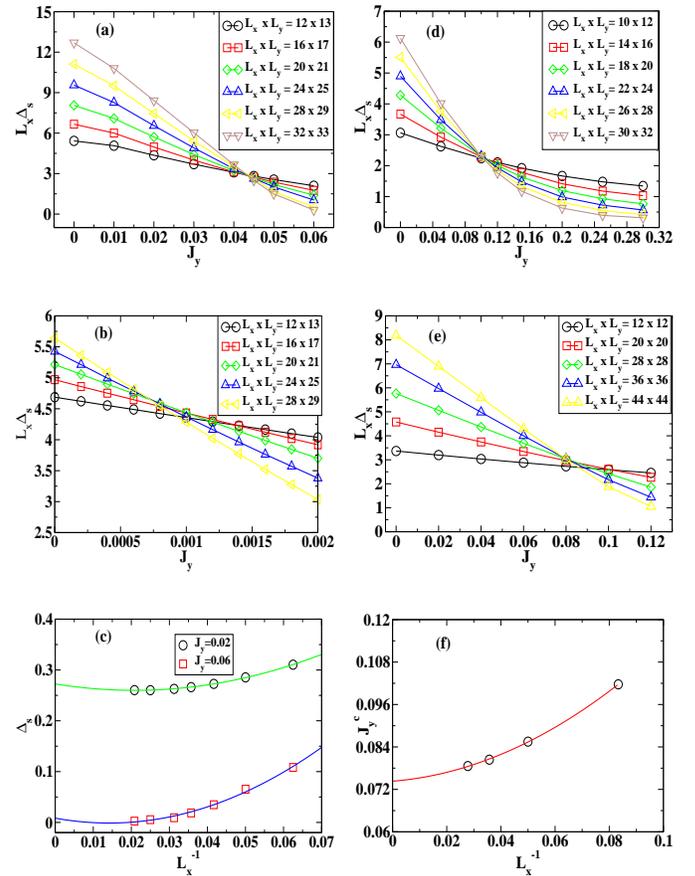

\begin{center}
$\begin{array}{c@{\hspace{0.5in}}c}
\vspace{0.4cm}
\includegraphics[width=4.25cm, height=3.5cm]{fig4.eps}
\hspace{0.25cm}
\includegraphics[width=4.25cm, height=3.5cm]{fig7.eps}
\end{array}$
$\begin{array}{c@{\hspace{0.5in}}c}
\vspace{0.4cm}
\includegraphics[width=4.25cm, height=3.5cm]{fig5.eps}
\hspace{0.25cm}
\includegraphics[width=4.25cm, height=3.5cm]{fig8.eps}
\end{array}$
$\begin{array}{c@{\hspace{0.5in}}c}
\includegraphics[width=4.25cm, height=3.5cm]{fig6.eps}
\hspace{0.25cm}
\includegraphics[width=4.25cm, height=3.5cm]{fig9.eps}
\end{array}$
\end{center}
\caption{ $\Delta_s \times L_x$ as function of $J_y$
 (a) for coupled $S=1$ chains, (b) for coupled $S=2$ chains;
(c) $\Delta_s$ as function of $L_x$ at two characteristics values
below ($J_y=0.02$) and above ($J_y=0.06$) the QCP for $S=1$
chains; (d) $S=\frac{1}{2}$ two-leg ladders at $J_r=0.4$; 
(e) and four-leg ladders with $J_r=1$; (f) $J_y^c$ as function of $L_x^{-1}$ 
in coupled four-leg ladders.}
\label{spins}
\end{figure}

In coupled chains with $S=1$, the ground state is known 
both in 1D and 2D. In 1D, the systems has a spin gap, the Haldane gap
$\Delta_{S=1}=0.41049$ \cite{golinelli}. The correlation length
$\xi_{S=1} \approx 6$ \cite{suzuki}. In 2D, we know from the rigorous result
 \cite{neves-perez,kubo} that the ground state has an LRO, thus it is 
gapless and $\xi_{S=1} = \infty$. Hence, there should be a QPT at some 
critical $J_y^c$ from a disordered to and ordered ground state.  QMC 
studies predict $J_y^c \approx 0.04$ \cite{kim-birgeneau} and 
$J_y^c=0.043648(8)$ \cite{matsumoto}. In the two-step study, we applied 
periodic boundary conditions in the $x$ direction and open boundary 
conditions in the $y$ direction. We show in Fig.\ref{spins}(a) the 
finite size behavior of the spin gap $\Delta_s$. It shows that in agreement 
with QMC, $J_y^c$ is located between $J_y=0.040$ and $J_y=0.045$. A more 
accurate DMRG estimate of $J_y^c$ can be obtained by locating the crossing 
point of consecutive $L_x \times L_y$ systems and extrapolating. But for 
$S=1$, an extrapolation was not necessary, for $L_x \times L_y \geq 16 \times 17$,
 the data converged around $J_y^c=0.043613$. 
We performed the same analysis for coupled $S=2$ chains. For a single chain 
$\Delta_{S=2}=0.0876$ \cite{wang} and $\xi_{S=2} \approx 49$ \cite{suzuki}. 
The scaled gaps are displayed in Fig.\ref{spins}(b), we find 
$J_y^c \approx 0.0007$. It is interesting to note that the values of $J_y^c$ 
 for $S=1$ and for $S=2$ are roughly consistent with the Schwinger bosons 
prediction, $J_y^c \times \xi_{1D}^2 \approx J_x$ \cite{azzouz-doucot}, where 
$\xi_{1D}$ is the 1D correlation length. We find 
$J_y^c \times \xi_{1D}^2 =1.5701$ for $S=1$, and 
$J_y^c \times \xi_{1D}^2 =1.5707$ for $S=2$. These results suggest that
$J_y^c \times \xi_{1D}^{2}= \frac{\pi}{2} J_x$. In Fig.\ref{spins}(c) 
we show $\Delta_s$ for two typical values in a $S=1$ system, 
$J_y=0.02 < J_y^c$ and $J_y=0.06 > J_y^c$. The extrapolated values are 
consistent with respectively gapped and gapless phases.

In $S=\frac{1}{2}$ two-leg ladders, if the coupling between the rungs
is $J_r=J_x=1$,  QMC studies \cite{matsumoto} predict that 
$J_y^c \approx 0.3$. We could not study the QPT because, the condition
$J_y^c \ll \Delta E$ is not fulfilled. If we reduce $J_r$ enough, the 
two-step DMRG becomes applicable. This is achieved when $J_r \alt 0.5$. 
For instance in Fig.\ref{spins}(d), we display the finite size behavior of 
$\Delta_s$ for $J_r=0.4$ and $L_x \times L_y$ ranging from $10 \times 12$ 
to $30 \times 32$. For these systems, we find $J_y^c=0.0993$. 
An alternative to reducing $J_r$ for $S=\frac{1}{2}$ systems in
order to study the QPT in $S=\frac{1}{2}$ systems is increasing the 
number of legs. When the number of legs increases, $\Delta_s$ on the
ladder decreases, the system is thus closer to criticality, therefore a 
smaller $J_y$ can induce a QPT. For the four-leg ladder with $J_r=1$, 
the QMC predicts $J_y^c \approx 0.07$ \cite{kim-birgeneau2}. 
In Fig.\ref{spins}(e), we show systems of four-leg ladders ranging from 
$12 \times 12$ to $44 \times 44$, it can be seen that $J_y^c \approx 0.08$. 
In Fig.\ref{spins}(f), we plot the crossing points of $12 \times 12$ and 
$20 \times 20$ to $36 \times 36$ and $44 \times 44$ systems respectively. 
This yields a better estimate of the QCP, $J_y^c=0.0742$.

 Let us now consider the anisotropic Hubbard model defined
by hopping parameters $t_x=1$, $t_y \ll t_x$ and a local
interaction $U$. The non-interacting single particle 
energies are, $\epsilon_{\bf k}= -t_x cos (k_xx)-t_y cos(k_yy)$ 
where ${\bf k}=(k_x,k_y)$. Since the Fermi surface is nested at the 
momentum ${\bf q_N}$, $\epsilon({\bf k}+{\bf q_N})=-\epsilon({\bf k})$.
 In the Hartree-Fock approximation, the consequence of nesting is 
that the  metallic state becomes unstable against the formation of a gap 
$\Delta_c^{HF} \approx \exp(\frac{-W}{4U})$ and spin-density wave 
LRO. In 1D, $\Delta_s^{1D}=0$ for all $U$ and the charge gap 
$\Delta_c^{1D} \approx \exp(\frac{-W}{4\sqrt{U}})$ for $U/W \ll 1$,
 and $\Delta_c^{1D} \approx U$ for $U/W \gg 1$. There is no LRO
in 1D, the gap opening cannot be explained in the Hartree-Fock
approximation. But it is generally believed that the Hartree-Fock
approximation is at least qualitatively correct when $t_y \neq 0$. 
 
However, there is a regime of the quasi-1D model where a simple physical
argument shows the failure of the Hartree-Fock approximation.
If $t_y \ll \Delta_c^{1D}$, interchain motion
is prohibited, the electrons are confined into the chains. 
For any $U>0$, the system would remain a Mott insulator.
There will be an LRO of the Heisenberg type because the spin 
degrees of freedom are gapless in 1D, and the small $t_y$ would 
yield an effective exchange, ${\tilde J}_y=t_x^2/\Delta_c^{1D}$. 
This regime cannot be described by the simple Hartree-Fock theory. 
This shows that a strong
coupling like behavior extends even for small $U$ in the confined
regime. This somewhat overlooked regime of the Hubbard model was discussed  
for two-coupled Hubbard chains \cite{lehur}. It is shown in Ref.\cite{lehur}
that if $t_y \ll \Delta_c^{1D}$, the systems is equivalent to the Heisenberg 
two-leg spin ladder. This is also implicit in the chain dynamical mean-field 
theory study which predicted that at half-filling, a finite
$t_y$ was necessary to deconfine the electrons in the transverse
direction. Hence, if we increase $t_y$ from the 1D Mott insulator at $t_y=0$ 
there are three possibilities: (i) the system remains a Mott insulator; 
(ii) there is a crossover from a Mott insulator towards a Slater insulator;
(iii) there is a QPT towards a metallic phase. 

\begin{figure}
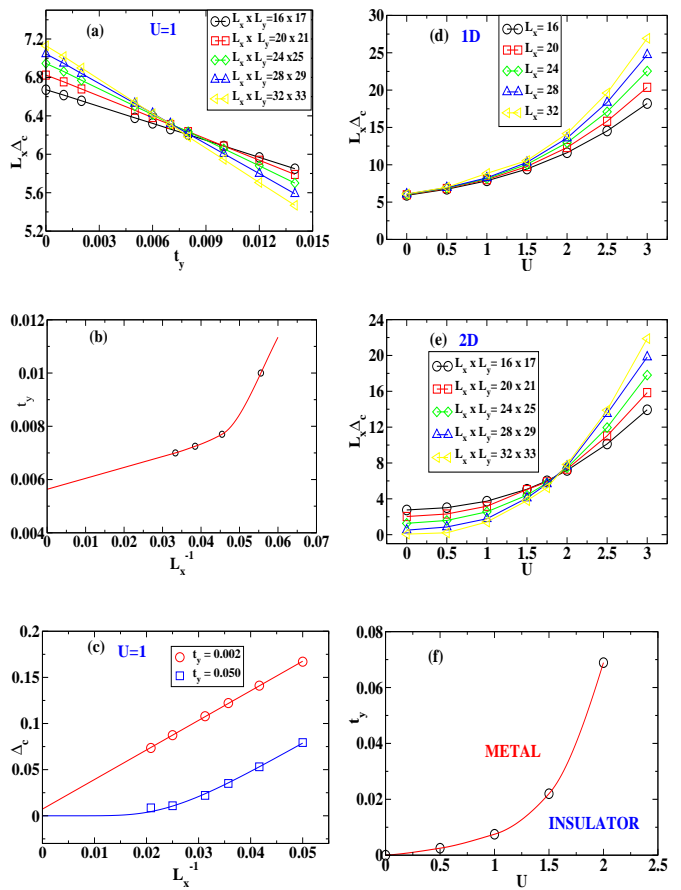

\begin{center}
$\begin{array}{c@{\hspace{0.5in}}c}
\vspace{0.4cm}
\includegraphics[width=4.25cm, height=3.5cm]{fig10.eps}
\hspace{0.25cm}
\includegraphics[width=4.25cm, height=3.5cm]{fig13.eps}
\end{array}$
$\begin{array}{c@{\hspace{0.5in}}c}
\vspace{0.5cm}
\includegraphics[width=4.25cm, height=3.5cm]{fig11.eps}
\hspace{0.25cm}
\includegraphics[width=4.25cm, height=3.5cm]{fig14.eps}
\end{array}$
$\begin{array}{c@{\hspace{0.5in}}c}
\includegraphics[width=4.25cm, height=3.5cm]{fig12.eps}
\hspace{0.25cm}
\includegraphics[width=4.25cm, height=3.5cm]{fig15.eps}
\end{array}$
\end{center}
\caption{(a) $\Delta_c \times L_x$ as function of $t_y$
    at $U=1$; (b) the deconfinement transition point $t_y^c$ at $U=1$ 
extrapolated from crossing points of two consecutive $L_x \times L_y$; 
(c) $\Delta_c$ as function of $1/L_x$ at $U=1$ for two characteristic 
values of $t_y$, below ($t_y=0.002$) and above ($t_y=0.05$) $t_y^c$;  
$\Delta_c \times L_x$ as function of $U$ in 1D (d) and 2D (e);
(f) Phase diagram of the quasi-1D Hubbard model in the $(U,t_y)$ plane.} 
\label{hubbard}
\end{figure}

The two-step DMRG results shown below are consistent with the case
(iii). In this study, we kept up to $m_1=384$ and $m_2=96$ respectively
during the first and second steps; the maximum truncation error was
less than $1 \times 10^{-5}$ in the first step and $1 \times 10^{-4}$ in
the second step for all the parameters investigated. The bulk of our 
calculations was on lattice sizes $L_x \times L_y$ ranging from 
$12 \times 13$ to $32 \times 33$, $U=0,~0.5,~1,~1.5,~2,~2.5,~3$, 
and at least $10$ different $t_y$ chosen from $t_y=0$ to 
$t_y \approx \Delta_c^{1D}$. In a few cases, $40 \times 41$ and 
$48 \times 49$ systems were also studied. In Fig.\ref{hubbard}(a) we show 
the scaled $\Delta_c$ as function of $t_y$ for $U=1$. The scaled gap displays
the typical behavior seen for spin systems. The data for different sizes
converge near $t_y=0.007$. The finite size behavior of $t_y^c$
(the crossing points of $L_x\Delta_c$ of consecutive systems) 
is shown in Fig.\ref{hubbard}(b). The extrapolation yields $t_y^c =0.0056$ 
for $U=1$. In Fig.\ref{hubbard}(c), we display $\Delta_c$ as function of 
$L_x$ at two typical values of $t_y$ above and below $t_y^c$. 
The extrapolated gap is in agreement with the qualitative behavior of 
$L_x\Delta_c$. Hence $\Delta_c$ displays a deconfinement transition from a 
1D Mott insulator to a 2D metallic phase. Using the chain-DMFT, Biermann 
and coworkers \cite{biermann} find that at $U=2.6$, the charge correlation 
exponent $K_{\rho}$ jumps from $0.02$ at $t_y=0.16$ to $1.01$ at $t_y^c=0.28$. 
This shows a deconfinement transition with $0.16 \alt t_y^c \alt 0.28$. 
We could not however see the $t_y$ induced QPT for  $U=2.6$. 
The relation $t_y^c \ll \Delta E$ was satisfied only 
for very small systems. The maximum value for which we could study
the $t_y$ induced QPT is $U=2$, we find $t_y^c=0.0689$. We obtain 
$t_y^c=0.18$ at $U=2.6$ by extrapolating from smaller values of $U$. 
Since in Ref.\cite{biermann} they were restricted to paramagnetic solutions, 
they could not rule out a possible AFM ground state of the Slater type. 

Once in  the deconfined regime,  we can induce a MIT
by increasing $U$. For this purpose, we set $t_y=0.05$. For this value
of $t_y$ we know from the calculations above that for any $U \alt 1.5$,
the system is in the metallic phase. We thus expect a MIT at some
$U$ between $U=1.5$ and $U=2$, because as seen above $t_y^c=0.0689$
at $U=2$. In Fig.\ref{hubbard}(d),(e), we show the scaled $\Delta_c$ 
as function of $U$ respectively in 1D and 2D. In the 2D case 
(Fig.\ref{hubbard}(e)), For $U \alt 1.875$, $L_x\Delta_c$ decays when we 
increase $L_x$ as for $U=0$ until it reaches $U_c \approx 1.8$ where, 
it starts to increase. This is to be contrasted to the 1D case shown in 
Fig.\ref{hubbard}(d) where there is no regime where $L_x\Delta_c$ decreases
when $L_x$ is increased, which implies $U_c=0$ as we know from the Lieb-Wu 
solution.

In Fig.\ref{hubbard}(f), we show the phase diagram of the quasi-1D Hubbard 
model. The deconfinement transition occurs for small, $U \alt 1$, at 
$t_y^c \approx \Delta_c^{1D}/4$. In a recent experiment, Pashkin and
coworkers \cite{pashkin} studied the infrared response of the quasi-1D Fabre 
salt $(TMTTF)_2PF_6$. Though this compound is nominally three-quarter filled, 
the presence of small dimerization renders it effectively half-filled. 
At ambient pressure, $(TMTTF)_2PF_6$ is a Mott insulator. When external
pressure is applied, the interchain transfer integral grows exponentially 
and the Mott gap rapidly decreases until a deconfinement transition is 
reached. The interchain transfer integral that induces the deconfinement
transition is approximately half of the Mott gap. This result is in 
reasonable agreement with our prediction.  

To conclude, let us comment on the implication of our result on the
isotropic case, $t_y=t_x$. For a given $U$, when $t_y > t_y^c$, the
system enters the metallic phase. It should remain in the 
metallic phase up to $t_y=t_x$, because, as soon as $t_y > t_y^c$, 
there is no other obvious process that will drive the system 
to a another phase when $t_y$ is further increased. 
In Ref.\cite{moukouri-jarrell} the DCA was applied to the isotropic 
2D Hubbard model at half-filling. It was found, for cluster 
sizes up to $N_c=64$, that down to $U=4$ the paramagnetic solutions 
remained gapped. Although no gap was found for $U < 4$, it was assumed 
that the gap would open for larger clusters which were not accessible. 
However, our result suggests that this assumption may not be true. 
We reexamined Ref.\cite{moukouri-jarrell} data, plotting $U_c(N_c)$,
where $U_c(N_c)$ is the estimated finite cluster QCP, 
 we find $U_c \approx 3.1$ when $N_c \rightarrow \infty$.

\begin{acknowledgments}
 We acknowledge a helpful correspondence with M. Jarrell, R.M. Scalettar, 
and A-.M.S. Tremblay. We  wish to thank A. Schiller for support.
This work was supported in part by a Shapira fellowship 
of the Israeli Ministry of Immigrant Absorption (S.M.), and by the Israel
Science Foundation through grant no. 1524/07.
\end{acknowledgments}


\begin{thebibliography}{99}
\bibitem{mott} N.F. Mott, Proc. Phys. Soc. (London) {\bf A62}, 416 (1949).
\bibitem{hubbard} J. Hubbard, Proc. Roy. Soc.(London) {\bf A277}, 237 (1964).
\bibitem{lieb-wu} E.H. Lieb and F.Y. Wu, Phys. Rev. Lett. {\bf 20}, 1445 
                  (1968).
\bibitem{metzner-vollhardt} W. Metzner and D. Vollhardt, Phys. Rev. Lett. 
                          {\bf 62}, 324 (1989).
\bibitem{georges-kotliar} A. Georges and G. Kotliar, Phys. Rev. {\bf B 45},
                          6479 (1992).
\bibitem{hirsch} J.E. Hirsch, Phys. Rev. {\bf B 31}, 4403 (1984).
\bibitem{varney} C.N Varney, C.-R. Lee, Z.J. Bai, S. Chiesa, M. Jarrell, and
                 R.T. Scalettar, Phys. Rev. {\bf B 80}, 075116 (2009).
\bibitem{moukouri-jarrell} S. Moukouri and M. Jarrell, Phys. Rev. Lett. 
                          {\bf 87}, 167010 (2001).
\bibitem{imada} Y.Z. Zhang and M. Imada, Phys. Rev. {\bf B 76},
                           045108 (2007).
\bibitem{park-haule} H. Park, K. Haule, and G. Kotliar, Phys. Rev. Lett.,
                           {\bf 101}, 186403 (2008).
\bibitem{moukouri} S. Moukouri, Phys. Rev. {\bf 70}, 014403 (2004).
\bibitem{biermann} S. Biermann, A. Georges, A. Lichtenstein, and T. Giamarchi,
                           Phys. Rev. Lett. {\bf 87}, 276405 (2001).
\bibitem{white} S.R. White, Phys. Rev. Lett. {\bf 69}, 2863 (1992).
\bibitem{barber} M.N. Barber in 'Phase Transitions and Critical Phenomena', 
                 edited by C. Domb and J. L. Lebowitz, Academic Press, London,
                 Vol. 8, p. 145 (1983).
\bibitem{neves-perez} E.J. Neves and J.F. Perez, Phys. Lett. {\bf A 114},
                           331 (1986).
\bibitem{kubo} K. Kubo and T. Kishi, Phys. Rev. Lett. {\bf 61}, 2585 (1988).
\bibitem{golinelli} O. Golinelli, Th. Jolicoeur, and R. Lacaze, Phys. Rev.
                          {\bf B 50}, 3037 (1994).
\bibitem{suzuki} N. Hatano and M. Suzuki, J. Phys. Soc. Jap. {\bf 62},
                           1346 (1993).
\bibitem{wang} X. Wang, S. Qin, Lu Yu, Phys. Rev. {\bf B 60},
                          14529 (1999). 
\bibitem{azzouz-doucot} M. Azzouz and B. Doucot, Phys. Rev. {\bf B 47},
                           8660 (1993).
\bibitem{kim-birgeneau} Y. J. Kim and R.J. Birgeneau, Phys. Rev. {\bf 62},
                           6378 (2000). 
\bibitem{matsumoto} M. Matsumoto, C. Yasuda, S. Todo, and H. Takayama,
                           Phys. Rev. {\bf 65}, 014407 (2001).
\bibitem{kim-birgeneau2} F.Y. Kim, R.J. Birgeneau, M.A. Kastner, Y.S. Lee, 
                          Y. Endoh, G. Shirane, and  K. Yamada, Phys. Rev.
                          {\bf B 60}, 3294 (1999).
\bibitem{lehur} K. Lehur, Phys. Rev. {\bf B 63}, 165110 (2001).
\bibitem{pashkin} A. Pashkin, M. Dressel, M. Hanfland, and C. A. Kuntscher,
                  Phys. Rev. {\bf B 81}, 125109 (2010).   
\end{thebibliography}
\end{document}